\documentclass[aip,amsmath,amssymb,reprint]{revtex4-1}
\usepackage{graphicx}
\usepackage{dcolumn}
\usepackage{bm}
\usepackage{soul}
\usepackage{color}
\usepackage{xcolor}
\usepackage{amsmath}
\usepackage{amssymb}
\usepackage{amsmath}
\usepackage{float}
\usepackage{natmove}
\usepackage{multirow}
\usepackage{setspace}
\usepackage{array}
\usepackage{booktabs}
\usepackage{rotating}
\usepackage{ulem}
\usepackage[colorlinks=true,linkcolor=blue,citecolor=blue,urlcolor=blue]{hyperref}
\usepackage{}
\usepackage{physics}
\usepackage{mathrsfs}

\begin{document}

%
%

\title{Narrowing of \textit{d} bands of FeCo layers intercalated under graphene }

%
%
%
\author{Daniela Pacil\`e*}
\affiliation{Dipartimento di Fisica, Universit$\grave{a}$ della Calabria, I-87036 Arcavacata di Rende, (Cs) Italy\looseness=-1}
\email{*Authors to whom correspondence should to addressed: daniela.pacile@fis.unical.it and claudiamaria.cardosopereira@nano.cnr.it}
\author{Claudia Cardoso*}
\affiliation{Centro S3, CNR-Istituto Nanoscienze, I-41125 Modena, Italy}
\author{Giulia Avvisati}
\affiliation{Dipartimento di Fisica, Universit$\grave{a}$ di Roma ``La Sapienza'', I-00185 Roma, Italy}

\author{Ivana Vobornik}
\affiliation{Istituto Officina dei Materiali (IOM)-CNR, Laboratorio TASC, Area Science Park, S.S.14, Km 163.5, 34149 Trieste, Italy}

\author{Carlo Mariani}
\affiliation{Dipartimento di Fisica, Universit$\grave{a}$ di Roma ``La Sapienza'', I-00185 Roma, Italy}
\author{Dario A. Leon}
\affiliation{Universit$\grave{a}$ di Modena e Reggio Emilia I-41125 Modena, Italy}
\affiliation{Centro S3, CNR-Istituto Nanoscienze, I-41125 Modena, Italy}
\author{Pietro Bonf\`a}
\affiliation{Università di Parma, Dipartimento di Scienze Matematiche, Fisiche e Informatiche, I-43124 Parma, Italy}
\affiliation{Centro S3, CNR-Istituto Nanoscienze, I-41125 Modena, Italy}
\author{Daniele Varsano}
\affiliation{Centro S3, CNR-Istituto Nanoscienze, I-41125 Modena, Italy}
\author{Andrea Ferretti}
\affiliation{Centro S3, CNR-Istituto Nanoscienze, I-41125 Modena, Italy}
\author{Maria Grazia Betti}
\affiliation{Dipartimento di Fisica, Universit$\grave{a}$ di Roma ``La Sapienza'', I-00185 Roma, Italy}
%
%
%
%
%
\begin{abstract}
We report on the electronic properties of an artificial system obtained by the intercalation of equiatomic FeCo layers under graphene grown on Ir(111). Upon intercalation, the FeCo film grows epitaxially on
Ir(111), resulting in a lattice-mismatched system. By performing Density Functional Theory calculations, we show that the intercalated FeCo layer leads to a pronounced corrugation of the graphene film. At the same time, the FeCo intercalated layers induce a clear transition from a nearly undisturbed to a strongly hybridized
graphene $\pi$-band, as measured by angle-resolved photoemission spectroscopy. A comparison of experimental results with the computed band structure and the projected density of states unveils a spin-selective hybridization between the $\pi$ band of graphene and FeCo-3\textit{d} states. Our results demonstrate that the reduced dimensionality, as well as the hybridization within the FeCo layers, induce a narrowing and a clear splitting of Fe 3\textit{d}-up and Fe 3\textit{d}-down spin bands of the confined FeCo layers with respect to bulk Fe and Co.
\end{abstract}

\maketitle

Ferromagnetic metals (FMs) and their alloys can be finely manipulated by changing their chemical composition~\cite{James_PRB_1999}, structural configuration~\cite{Lu_APL_2007}, and by reducing the dimensionality~\cite{Getzlaff_APA_2006, Moulas_PRB_2008}. Modified symmetry and scaled dimension, indeed, may induce in FMs higher magnetic moments and larger uniaxial magnetic anisotropy energy (MAE) with respect to their 3D counterparts~\cite{Gambardella_Science_2003, Nonas_PRL_2001, Moulas_PRB_2008, Gambardella_Nature_2002, Avvisati_ASS_2020}. The enhanced magnetism in nanostructures, can be used e.g, for engineering spintronic devices~\cite{Sethulakshmi_MaterTod_2019}, high-density magnetic storage~\cite{Terris_JPD_2005}, and permanent magnets~\cite{ Bhatti_MaterTod_2017}.  

Among all iron-based alloys, including also pure iron, equiatomic FeCo exhibits the highest Curie temperature and magnetic moment driven by an almost filled Fe majority spin band~\cite{Bardos_JAP_1969, DiFabrizio_PRB_1989, Paduani_JAP_1999, IronAlloys}. On the other hand, due to its cubic symmetry, the FeCo alloy shows also a low MAE~\cite{Sundar_IntMagRev_2005}. Indeed, modifying dimensionality or symmetry has been shown to be a viable route in order to improve the magnetic response of FeCo alloys~\cite{Hasegawa_SciRep_2017, Avvisati_ASS_2020}, although clustering and granularity may hinder a controlled growth of artificial systems~\cite{Einax_RevModPhys_2013}. Intercalation of metals underneath graphene (Gr) has led to the formation of atomically smooth metallic layers~\cite{BATZILL201283, Dedkov_2011, Dedkov_2015, Decker_PRB_2013,Pacile_PRB_2014, Pacile2013, Vita_PRB_2014,Dedkov_2017, Cardoso2021, Avvisati_JPCC_2017, YANG2020100652}, where Gr on top is also a protective membrane against contaminants~\cite{Weatherup2015JACS,Bazarnik_SurfSci_2015}. Although several metals have been intercalated under Gr, the combined intercalation of Fe and Co, to form an equiatomic alloy is more challenging, as segregation without intermixing may be dominant~\cite{Bazarnik_SurfSci_2015}. 
Recently, we have succeeded in the growth of homogeneous and smooth Fe$_{x}$Co$_{1-x}$ layers underneath Gr/Ir(111), co-depositing Fe and Co at the same evaporation rate and keeping the substrate temperature at about 500 K, to avoid any Fe-Ir and Co-Ir alloying formation and clustering. By performing X-ray Magnetic Circular Dichroism (XMCD) measurements, we showed that Fe and Co are well intermixed and magnetically coupled underneath Gr, leading to an enhancement of the magnetic moments with respect to pure Fe and Co films, and exhibiting a strong ferromagnetic exchange coupling between the two species~\cite{Avvisati_ASS_2020}. 

In the present work, we investigate the electronic and structural properties of the equiatomic FeCo layers (about 1-2 monolayers of thickness) grown epitaxially on Ir(111) underneath a Gr cover, by means of Angular Resolved Photoemission Spectroscopy (ARPES) and Density Functional Theory (DFT). We show that the reduced dimensionality in the artificially confined system leads to a narrowing and a redistribution of majority and minority 3\textit{d} states with respect to the pure species, enhancing also the splitting between Fe 3\textit{d}-up and Fe 3\textit{d}-down spin bands. Our description of the electronic structure in equiatomic FeCo layers sheds light on the enhanced magnetism previously demonstrated by XMCD~\cite{Avvisati_ASS_2020} and provides a useful insight in the engineering of low dimensional FM alloys. 

The Ir(111) surface was cleaned by cycles of sputtering (Ar$^{+}$, 1.5--2.0 keV) and annealing at about 1300~K.  The Gr sheet is obtained by exposing the clean surface to 5$\times$10$^{-8}$--2$\times$10$^{-7}$ mbar of C$_2$H$_4$ and annealing at 1300-1320 K. Metallic Fe and Co were simultaneously deposited, at previously calibrated same evaporation rates (about 0.3 \AA/min) to deposit nominal 3 monolayers (ML) of FeCo, on the Gr/Ir(111) substrate kept in the range of 500-530 K. As recently reported~\cite{Avvisati_ASS_2020}, core level photoemission and absorption spectroscopy unveil homogeneous intercalation of equiatomic FeCo layers, without any alloy formation with the underlying substrate ~\cite{Avvisati_ASS_2020}{(as shown in the Supplementary Material)}, at variance e.g. with recent observations for Mn intercalation on Gr/Ir~\cite{voloshina2021}.

 \begin{figure}
\includegraphics[width =0.47\textwidth]{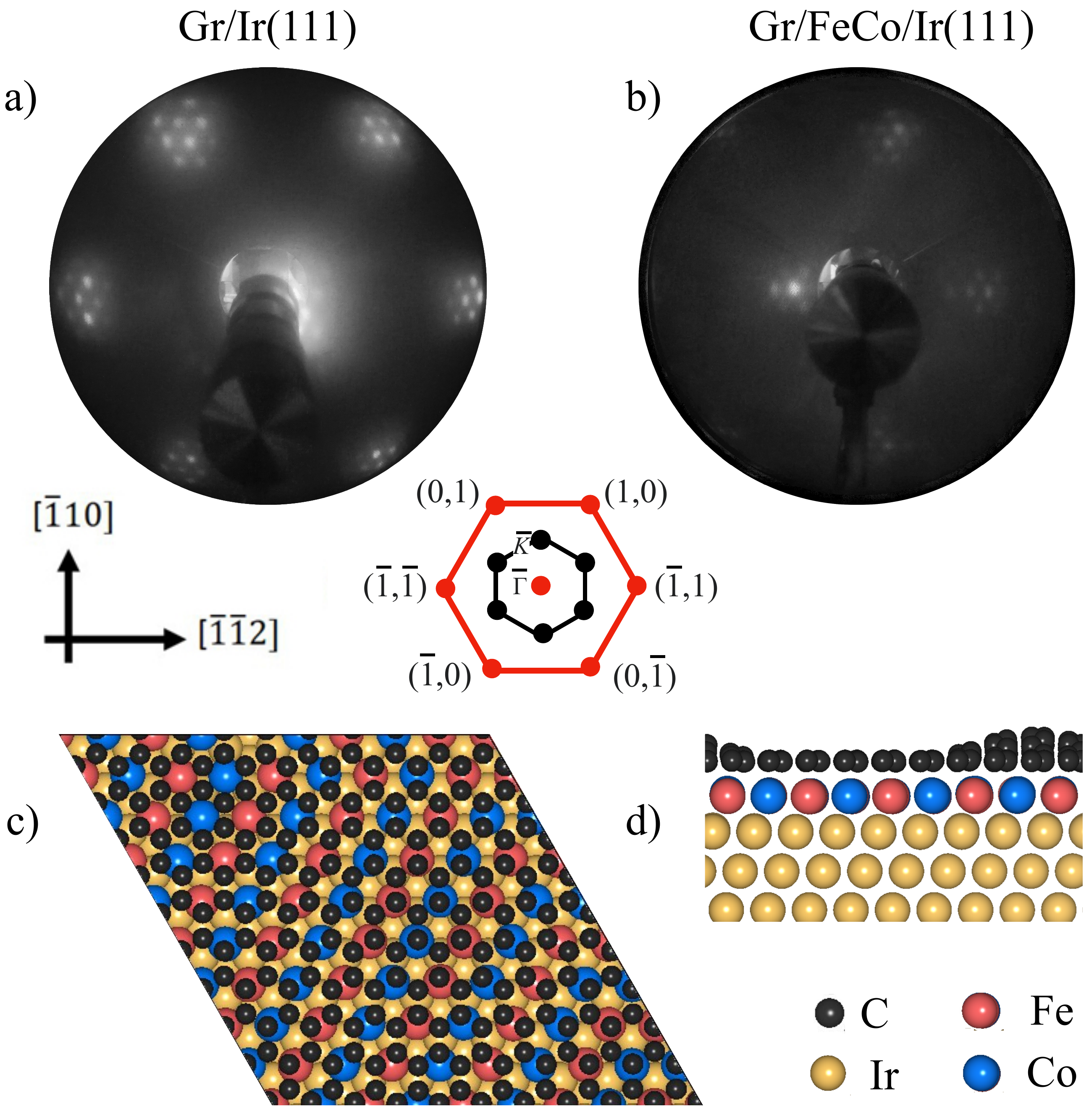}
\caption{\label{fig:LEED} Low energy electron diffraction (LEED) patterns (electron beam energy 90 eV) of (a) Gr/Ir(111) and (b) Gr/FeCo/Ir(111), taken slightly off-normal. In the middle: sketch of the diffraction pattern and Brillouin Zone (BZ) of Ir(111), referred to (a-b) panels. (c) Atomic structure as deduced by DFT, top view of the moir\'e pattern of Gr/FeCo/Ir(111) with C atoms represented in gray, Fe in red, Co in blue and Ir in cream; (d) side view of the Gr/FeCo/Ir(111) structure showing the Gr corrugation, using the same colors described in the previous panel. 
}
\end{figure} 

The sample preparation and ARPES experiments were carried out at the APE beamline of the Elettra Synchrotron Light Source (Trieste, Italy). The photoelectrons, excited with photon energy of 86 eV, were collected with a VG Scienta DA30 electron energy analyzer, which operates in deflection mode and allows for detailed $\mathbf{k}$-space mapping at fixed sample geometry.

\begin{figure}
\includegraphics[width=\linewidth]{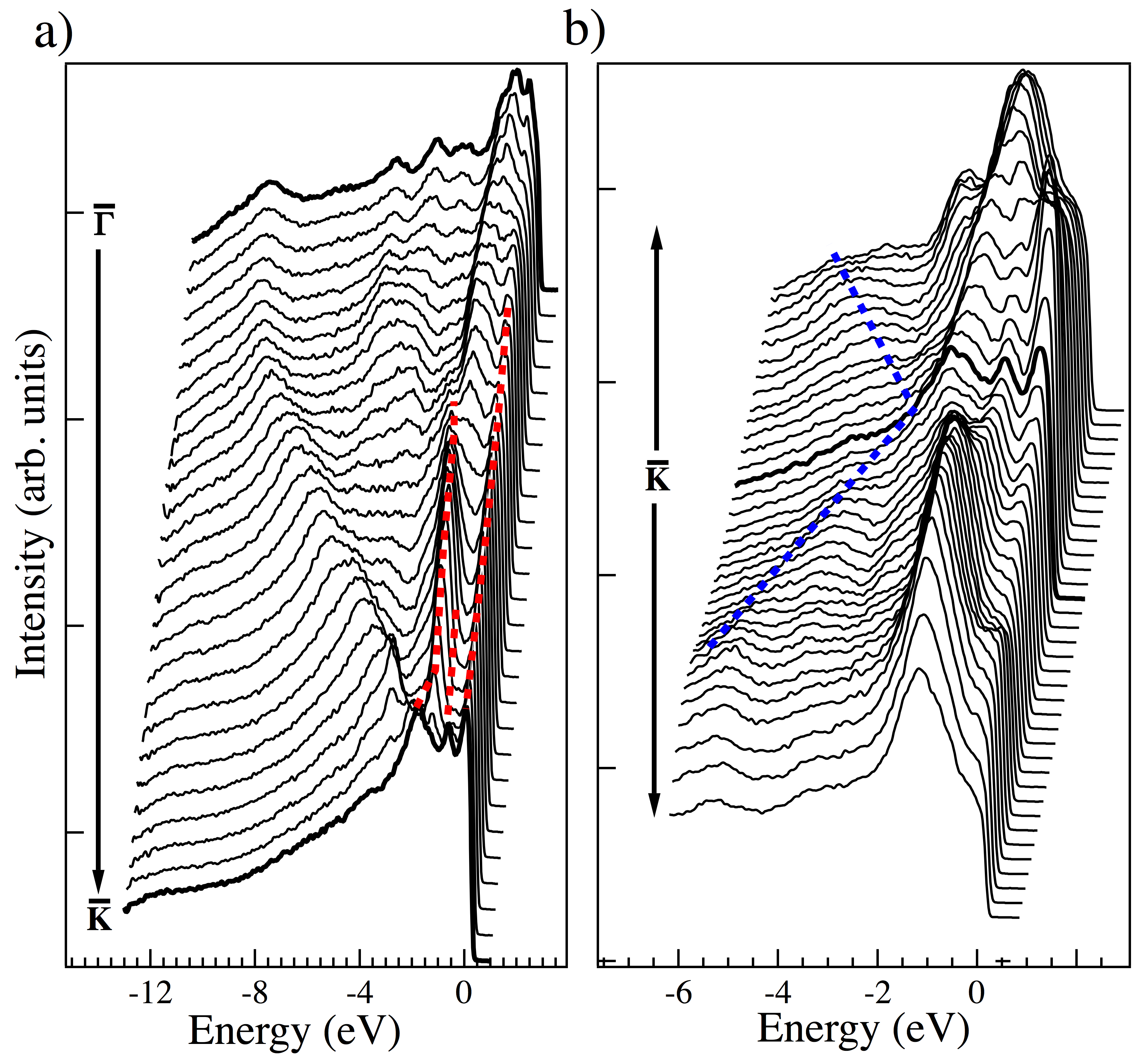}
\caption{\label{fig:arpes1} (a) Selected EDCs of Gr/FeCo/Ir(111), extracted from data of Fig.~\ref{fig:band_arpes}(a), acquired with photon energy of 86 eV and taken along the $\bar\Gamma$-$\bar{\mathrm K}$ direction. Red dashed lines highlight states originating from the FeCo interface. (b) Selected EDCs taken along the direction perpendicular to $\bar\Gamma$-$\bar{\mathrm K}$. Blue dashed lines highlight the $\pi$ band of Gr. The $\bar{\mathrm K}$-point has been extracted at electron kinetic energy of about 80 eV. 
}
\end{figure} 

Density functional theory (DFT) simulations were carried out using the plane wave implementation of the \textsc{Quantum ESPRESSO}~\cite{Giannozzi2009,Giannozzi2017} distribution.
We used the local density approximation (LDA) for the exchange-correlation potential, according to the Perdew-Zunger parametrization~\cite{Perdew_PRB_1981}. Since LDA is known to underestimate the values of the orbital magnetic moments in transition metals~\cite{Truhlar09}, we have adopted a DFT+U scheme~\cite{cococcioni2005linear}, with a Hubbard U parameter of 2~eV for Fe and 4~eV for Co. The choice of the values is explained in detail in the SI.
Similarly to our previous works~\cite{Avvisati_NanoLetters_2018, CalloniJCP2020, Cardoso2021}, we simulated the Gr/1ML-FeCo/Ir(111) interface considering the complete moir\'e induced periodicity by using a 9$\times$9 supercell of Ir(111), corresponding to a 10$\times$10 supercell of pristine Gr. More details about the structure can be found in the Supplementary Material.

\begin{figure*}
\includegraphics[width =0.99\textwidth]{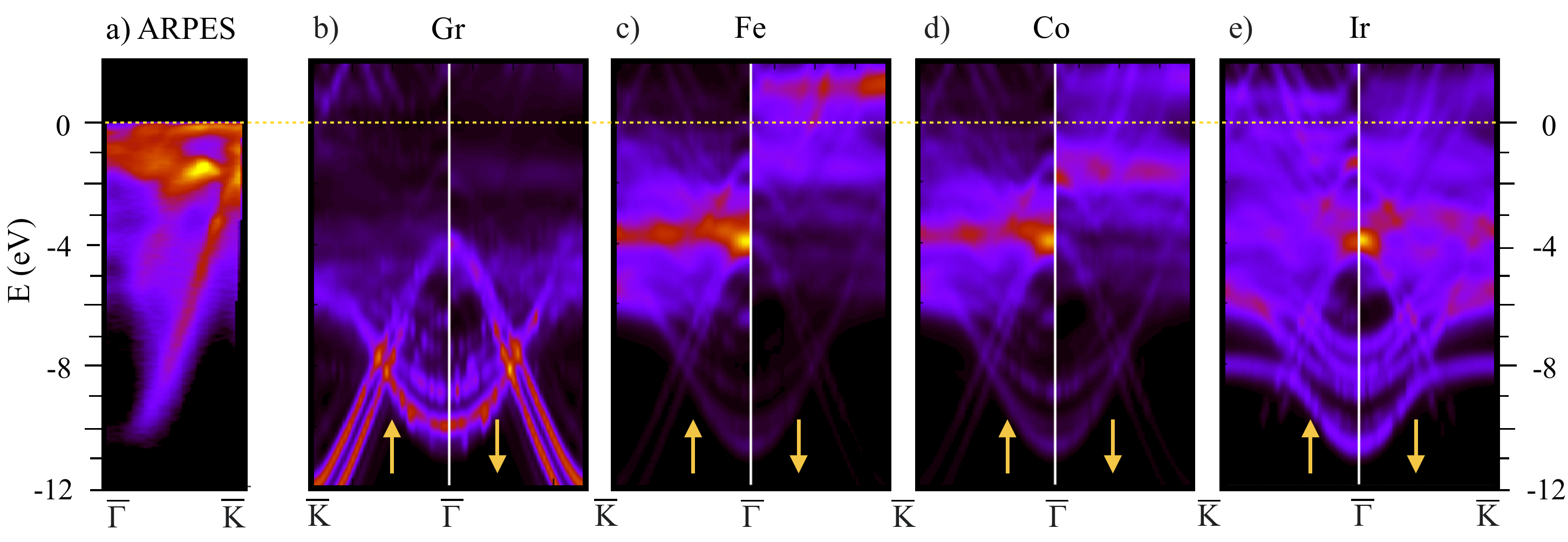}
\caption{\label{fig:band_arpes}
Comparison between ARPES measurements of Gr/FeCo/Ir(111) taken along the $\bar\Gamma$-$\bar{\mathrm K}$ direction (a), and the band structure computed for the 10$\times$10 supercell of Gr/1ML-FeCo/Ir(111), mapped into the graphene 1$\times$1  BZ and projected  into the atomic orbitals of the different atomic species (b-e).  
}
\end{figure*}

In order to obtain a simpler representation of the Gr/FeCo/Ir(111) band structure, we have applied an unfolding procedure~\cite{Popescu2012,unfold-x} in which the band structure computed for the 10$\times$10 supercell is mapped into the Gr 1$\times$1  Brillouin Zone (BZ) by using the {\tt unfold-x} code~\cite{unfold-x}. 
This approach has been further extended to include atomic projections on L\"odwin orthogonalized atomic orbitals as provided by the \textsc{Quantum ESPRESSO} package~\cite{Giannozzi2009,Giannozzi2017}.
 

Figure~\ref{fig:LEED} shows the LEED patterns of Gr/Ir(111), taken with primary energy of about 90~eV, before (a) and after (b) the intercalation of nominal 3ML of FeCo. The moir\'e superstructure characteristic of the corrugated Gr/Ir(111) system is maintained after FeCo intercalation, attesting that the equiatomic FeCo layers arrange pseudomorphycally on the hexagonal Ir(111) surface. The moir\'e pattern is only slightly smeared out at increasing amount of intercalated atoms, with an intensity reduction of the extra spots. The persistence of multiple reflection spots suggests that the thickness of the intercalated FeCo layer is within 1 and 2 ML, as deduced by comparison with the LEED patterns of pure Co~\cite{Pacile_PRB_2014} and Fe~\cite{Cardoso2021} intercalated systems. 

The Gr/FeCo/Ir(111) interface was modeled at the DFT level as described previously. The calculations indicate a large corrugation of the Gr layer (1.40~\AA), with a Gr-FeCo interplanar distance of 1.90~\AA\ and  3.30~\AA\ at valleys and hills, respectively, as shown in Fig.~\ref{fig:LEED}(c-d). The corrugation of the Gr/FeCo/Ir(111) and the energy landscape of the stable structural configuration are similar to those computed for the pure intercalated systems Gr/Fe/Ir(111) (1.3~\AA)~\cite{Cardoso2021} and Gr/Co/Ir(111) (1.4~\AA)~\cite{Avvisati_NanoLetters_2018,Decker_PRB_2013}. 
It is worth noticing that we do not include van der Waals interactions in our approach since LDA has shown to give a good description of graphene corrugation and distance for Gr/Fe/Ir(111)~\cite{Cardoso2021}, providing essentially the same scenario as DFT calculations done at the PBE level including van der Waals interactions~\cite{Decker_JPhys_2014}, which only find slightly larger Gr-Fe distances.

\begin{figure}
\includegraphics[width =0.47\textwidth]{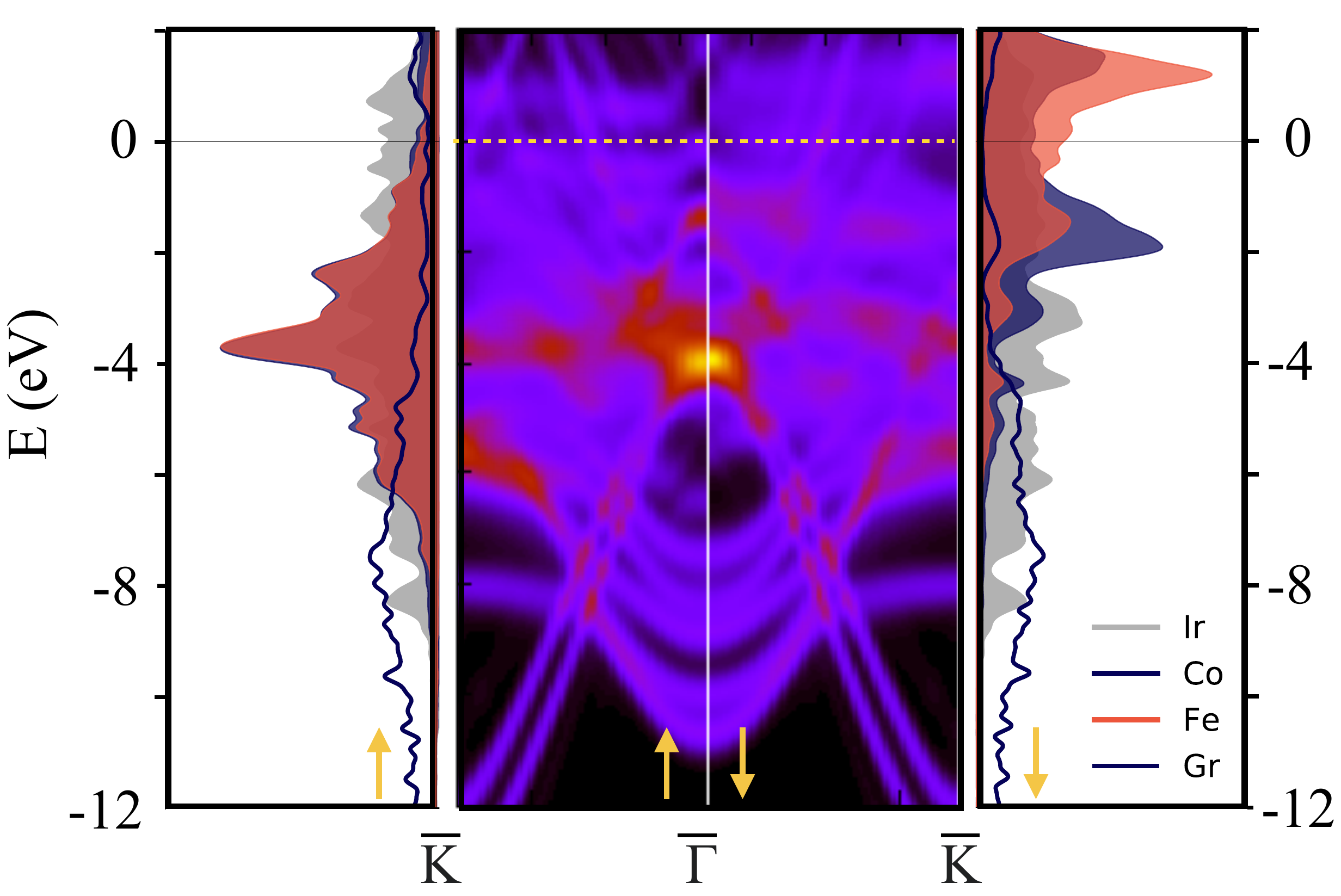}
\caption{
 Central panel: majority and minority spin band structure computed within DFT for  Gr/1ML-FeCo/Ir(111). Left (right) panel: majority (minority) spin DOS computed within DFT and projected on C, Fe, Co and Ir atomic orbitals.
\label{fig:DOS_bands}  
}
\end{figure}

To shed light on the interaction mechanism between Gr and the FeCo interface, we show angular resolved photoemission measurements and compare them with the band structure computed using Kohn-Sham DFT.
Figure~\ref{fig:arpes1}(a) shows ARPES results of Gr/FeCo/Ir(111) taken along $\bar\Gamma$-$\bar{\mathrm K}$. The effect of Fe and Co states on the $\pi$ and $\sigma$ bands of Gr can be discussed by comparing the system under study with pure Gr/Co and Gr/Fe (either supported on Ir(111) or not), as reported in the literature~\cite{Pacile_PRB_2014, Varykhalov_PRX_2012, Brede2016, Usachov_2015, Cardoso2021}. At the $\bar\Gamma$-point (Fig.~\ref{fig:arpes1}a)
the bottom of the $\pi$ band and the top of the $\sigma$ bands are shifted to about 10.2 eV and 5.4 eV of binding energy (BE), respectively, in agreement with previous results for pure Fe and Co intercalated systems~\cite{Cardoso2021, Pacile_PRB_2014}.  Moreover, 
ARPES data extracted along the direction perpendicular to $\bar\Gamma$-$\bar{\mathrm K}$ (Fig.~\ref{fig:arpes1}b) shows 
the maximum of the Gr $\pi$ band at about 2.1~eV of BE. Other pure Gr/FM systems exhibit a $\pi$ band maximum between 2.5 and 3.0 eV, depending both on the metal and
on the number of intercalated layers \cite{Pacile_PRB_2014, Varykhalov_PRX_2012, Brede2016, Usachov_2015, Cardoso2021, massimi2014}. The Gr/Co system, e.g., leads to a crossing point between $\pi$ bands located at 2.8 eV of BE~\cite{Pacile_PRB_2014}. For Gr/FeCo/Ir, the lower BE value of the $\pi$ band maximum suggests a reduced width of the FeCo-\textit{d} states and thus a reduced hybridization with the $\pi$ band.

The region in proximity of the $\bar{\mathrm K}$-point exhibits several features directly related to the mixed Fe-Co interface with Gr. Specifically, in Fig.~\ref{fig:arpes1}(a) we highlight with red dashed lines three slightly dispersing states, crossing the $\bar{\mathrm K}$-point at about -1.8 eV, -0.6 eV and at the Fermi level ($E_\text{F}$). It is worth noticing that the appearance of these well-defined localized bands within $\sim$2~eV from $E_\text{F}$ and their dispersive behavior provide further evidence of a uniform FeCo alloy. 
The bands at about 4 eV of BE, less dispersive towards the $\bar{\mathrm K}$ point and resonant with the $\pi$ band of Gr, can be attributed to the FeCo interlayer. As discussed in the following, these states are due to the majority 3\textit{d} states of both Co and Fe atoms, while the states close to $E_\text{F}$ can be associated to the Fe and Co 3\textit{d} spin minority bands.  

The experimental evidence of narrow bands associated to FeCo in the energy region of $E_\text{F}$ and the hybridization of $\pi$ states of graphene is confirmed by the DFT calculations.  The spin-resolved electronic structure of Gr/1ML-FeCo/Ir(111), shown in Fig.~\ref{fig:band_arpes} (b-e), is projected on atomic orbitals of different species and mapped into the 1$\times$1 Gr BZ along the $\bar\Gamma$-$\bar{\mathrm K}$ direction. With respect to free-standing Gr, the main effect on the $\pi$ and $\sigma$ bands is a non-rigid shift towards higher BE (Fig.~\ref{fig:band_arpes}b), as also found in the literature for pure intercalated systems~\cite{Pacile_PRB_2014, Varykhalov_PRX_2012, Brede2016, Usachov_2015, Cardoso2021}. In addition, our calculations clearly show a significant interaction of the Gr $\pi$-band with the FeCo-\textit{d} majority states, that is stronger than with the minority bands. This is especially evident in the wide 0-5 eV energy range below $E_\text{F}$(Fig.~\ref{fig:band_arpes}a). The spin-down $\pi$ band of Gr preserves its linear dispersion up to about -2 eV, where it strongly overlaps with the FeCo minority states {(Fig.~\ref{fig:band_arpes}b-d)}, and their interaction pushes the $\pi$ band maximum down, in agreement with our ARPES measurements. Therefore, in line with a recent paper of Gr grown on a Mn$_5$Ge$_3$ FM interface~\cite{voloshina2021}, the theoretical predictions demonstrate that the interaction of Gr with the FeCo layer gives rise to a downshifted Gr $\pi$ band which is mainly minority-spin polarized. 

This picture is further confirmed and summarized by the comparison of the total band structure  (central panel of  Fig.~\ref{fig:DOS_bands}), with the spin DOS projected on C, Fe, Co and Ir atomic orbitals (left-right panels of Fig.~\ref{fig:DOS_bands}), computed within DFT for Gr/1ML-FeCo/Ir(111). From the pDOS we can see that the Fe and Co 3\textit{d} states, particularly in the majority spin channel, almost coincide (see the Supplementary Material for a detailed analysis, including the role of DFT+U corrections). The majority spin channel is almost filled, with an intense peak centered at about -4~eV, where the spin-up Dirac cone is strongly hybridized and almost quenched. In the minority spin channel, the main superimposition of Fe and Co pDOS peaks is seen at about -2~eV and at the Fermi level, where indeed the transition metal states appear in our ARPES measurements. 

The density of states and the spin electronic bands for bulk FeCo alloys have been deeply investigated and discussed in the literature~\cite{Victora1984, Moulas_PRB_2008, Pizzini_PRB_1994}.  The strong hybridization of Fe and Co in the bulk alloy produces an increase of the Fe exchange splitting and a saturation of the magnetic moments~\cite{Pizzini_PRB_1994, Victora1984}. 
More specifically, it has been shown that at about 30\% Fe content the atomic magnetic moments become almost independent on the alloy composition, due to a redistribution of 3$d$ minority electrons to the 3$d$ majority states at the Fe sites.
Furthermore, the reduced dimensionality and number of nearest neighbors at the FeCo interface also influence the $d$ bandwidth and the spin splitting of the states~\cite{Pizzini_PRB_1994, Victora1984, Moulas_PRB_2008}. 

Despite a number of differences with the Gr/FeCo/Ir interfaces studied here, such as the interaction with Gr or with the Ir surface,  some of the features of FeCo alloys subsist. In particular, we observe a larger spin-splitting between majority and minority spin channels, and a narrowing bandwidth of the 3$d$ states for this confined FeCo layer with respect to bulk Fe and FeCo~\cite{Victora1984}. A detailed comparison between the Gr/FeCo/Ir(111) interface and the bulk systems is presented in the SI.
From the DFT simulations, the spin-resolved DOS of Fe and Co show almost completely filled 3$d$-states in the spin-up channel, with about 
4.76(4.70) electrons for Fe(Co), while about 2.08(3.09) electrons are localized in the 3$d$ spin-down channel.
Though the energy position of the FeCo 3$d$ states depends on the values of U chosen for the Hubbard correction, there is always a significant overlap between the Fe and Co states, while the Gr states are almost insensitive to different values of U, as described in the Supplementary Material.

Therefore, the results shown in Fig.~\ref{fig:DOS_bands} demonstrate that the reduced symmetry and the structural configuration of the intercalated equiatomic FeCo layer induce a clear splitting of Fe 3\textit{d}-up and down bands, with a narrower bandwidth with respect to Fe bulk. 

This picture is consistent with previous XMCD measurements on Gr/FeCo/Ir(111), showing a sensible enhancement of the magnetic moments~\cite{Avvisati_ASS_2020} with a strong ferromagnetic coupling between Fe and Co.

In conclusion, within the present study we characterize the electronic and structural properties of few FeCo layers, embedded between Gr and Ir(111). By means of ARPES we show that the interaction of the $\pi$ band of Gr with FeCo-3\textit{d} states pushes the strongly hybridized Dirac cone towards higher BE in analogy with pure intercalated systems. However, in proximity of $E_\text{F}$ we identify new localized states associated to the FeCo alloy. By means of DFT we show that the homogeneous intermixing of the two FMs and the artificial structural phase, where the FeCo layer is stretched to the Ir(111) lattice constant, leads to a narrowing and to an enhanced spin splitting of the 3\textit{d} states with respect to pure bulk systems. From an experimental point of view, we have succeeded in inducing such narrowing and enhanced spin splitting within a homogeneous 2D FeCo system. Therefore, our study provides a new artificial 2D FeCo system protected from contaminants through the Gr layer, inducing an enhanced magnetic response, which can be further engineered for integration in magnetic devices. 

See the Supplementary Material for: selected core levels measurements; additional calculations of projected DOS with different values of Hubbard U parameter; projected DOS for $10\times 10$ Gr/FeCo/Ir(111), $10\times 10$ Gr/Fe/Ir(111), $10\times 10$ Gr/Co/Ir(111) and DOS for corresponding FM bulk systems.

This work was partially supported by the MaX -- MAterials design at the eXascale -- Centre of Excellence, funded by the European Union program H2020-INFRAEDI-2018-1 (Grant No. 824143), by PRIN FERMAT (2017KFY7XF) from Italian Ministry MIUR and by Sapienza Ateneo funds.
Computational time on the Marconi100 machine at CINECA was provided by the Italian ISCRA program. This work has been partly performed in the framework of the nanoscience foundry and fine analysis (NFFA-MIUR Italy Progetti Internazionali) facility. The authors thank A. Vegliante for his contribution to the experiment.

%

The data that support the findings of this study are available from the corresponding author upon reasonable request.
%
%

\end{document}